\documentclass[12pt]{iopart}
\usepackage[utf8]{inputenc}
\usepackage{amsfonts}
\usepackage{amssymb}
\usepackage{wasysym}
\usepackage{graphicx}
\begin{document}
\title{Positron extraction to an electromagnetic field free region}
\author{D A Cooke$^1$, G Barandun$^1$, S Vergani$^1$, B Brown$^2$, A Rubbia$^1$ and P Crivelli$^1$}
\address{$^1$ ETH Zurich, Institute for Particle Physics, Otto-Stern-Weg 5, CH-8093, Zurich, Switzerland}
\address{$^2$ Physics Department, Marquette University, 1250 W. Wisconsin Avenue, Milwaukee, WI 53233, USA}
\ead{dcooke@phys.ethz.ch}

\begin{abstract}
We describe a scheme for high efficiency ($\sim 90\%$) extraction of 50 ns positron bunches from a buffer gas trap in an electromagnetic field free region. The positrons are time bunched to approximately 1 ns (FWHM) and focussed to less than 1 mm ($\sigma$). The target is kept at ground potential which is an advantage for many applications. The results compare well with SIMION simulations.
\end{abstract}

\section{Introduction}
The positron buffer gas trap allows the conversion of a `traditional' positron beam (typically generated using a radioactive source followed by a moderator) into a high-quality, pulsed beam with very low energy spread and small radius. Based around confinement of positrons in a Penning--Malmberg trap \cite{Surko1989}, the basic principle is that positrons pass over a potential barrier at the entrance to the trap, into a region with a high ($\sim 1\times10^{-3}$ mbar) pressure of a buffer gas (typically N$_2$). Radial confinement is provided by the magnetic field, and the potentials of the trap electrodes are such that, following an inelastic collision with a gas molecule, a positron no longer possesses sufficient energy to surmount the entrance barrier, and a similar barrier exists at the exit of the trap, meaning the positron is trapped. The potential well structure of the trap means that as the positrons lose energy through collisions with the gas they are eventually (on a millisecond timescale) confined to the end of the trap. The pressure at this second stage of the trap is considerably lower than at the inlet and first stage, so the positrons can be stored at this stage for some time while more positrons enter the trap. When required, the exit electrode can be lowered, releasing a pulse of positrons. This basic method can be improved upon by applying a rotating electric field to the positron cloud in the second stage \cite{Huang1997}. This compresses the cloud and counteracts a major loss mechanism and limitation on the second stage positron lifetime: radial transport caused by gas molecule collisions. The rotating wall method, as it is known, also heats the positron cloud, so a cooling mechanism must be employed. If the trap magnetic field is high enough, this can be cyclotron cooling, though it is less technically challenging to use a cooling gas (such as CF$_4$ or SF$_6$). Using this method, positron pulses with thermal energy distributions and radial dimensions around 1 mm can be achieved (see e.g. \cite{Danielson2015} and references therein).\par

Since their development \cite{Murphy1992}, buffer gas traps have found applications in many areas of positron physics, from scattering experiments (e.g. \cite{Barnes2004}) to antihydrogen formation \cite{Amole2014,Enomoto2010,Doser2012,Perez2012}. For many applications, however, the significant magnetic field required in the trapping region is an obstacle, as the experiment should be performed in a magnetic field free region. Scattering experiments which hope to record doubly or triply differential cross-sections, such as a positron reaction microscope \cite{Slaughter2009,Williams2010}, require an electrostatic positron beam, as do experiments using brightness enhancement methods relying on electrostatic focussing of a positron beam \cite{Cassidy2007}. For spectroscopic experiments on Ps \cite{Cooke2015}, there is a clear advantage to performing in a completely field free region (i.e. with no electric or magnetic field). To achieve this, the target region should be ideally at 0 V relative to the vacuum chamber, which is also a useful feature for other experiments where the target should be, for example, heated or cooled. Additionally, some scattering experiments which rely on a scattering cell held at a potential in order to define the impact energy \cite{Sullivan2008,Marler2005} may benefit from using grounded cells with the beam energy fixed earlier in the apparatus to avoid possible energy distortions introduced by cell fringe fields.\par

A number of different solutions exist in literature for the extraction of particles to zero magnetic field. The first \cite{Gerola1995} consists in placing at the extraction point an iron magnetic shield with a grid in the middle made of rings and bars. The idea is to break the cylindrical symmetry and apply the Busch's theorem locally, in order to reduce the transverse momentum and energy of the positrons. The second method consists of a remoderator made out of nickel \cite{Fujinami2008} or tungsten \cite{Oshima2008} placed after the extraction point. The remoderators have a thickness of 150 and 200 nm, and before and after them focus lenses are placed. The third method \cite{Weber2010} consists in placing a mu-metal shield after the extraction point followed by an einzel lens system. In this configuration, due to the non-adiabatic extraction, the beam size should remain the same. A summary of the efficiencies and extracted beam sizes of these methods is shown in Table \ref{tab:methods}. We decided to implement the last method, because production of the magnetic grid required for the first method was technically challenging, and the remoderation method has an intrinsic efficiency too low for our purposes. We made essentially two variations to the original solution. The first one was a further stage, an `elevator', put between the trap and the extraction point. With this device, which consists of a cylindrical electrode pulsed to high voltage when the positron bunch is inside it, we were able to get positrons at higher energies (up to 5 keV), which increases the extraction efficiency. This allows us to maintain the target at ground, as distinct from previous experiments, which is useful for our experiment and several other applications (as discussed above). 

\begin{table}
\centering
\begin{tabular}{l|ll}
method           & final beam size (sigma) 	& efficiency \\\hline
magnetic grid    & 3.7 mm					& 80\%             \\
remoderator (Ni) & 25 $\mu$ m				& 4.2\%            \\
remoderator (W)  & 38 $\mu$ m				& 10\%             \\
mu-metal shield  & 3.5 mm					& 55\%             
\end{tabular}
\caption{Different beam sizes obtained using different methods}
\label{tab:methods}
\end{table}

\section{Experimental technique}
Low energy positrons are continuously produced using a $^{22}$Na source coupled to a moderator comprising of a stack of annealed tungsten meshes. A two-stage N$_2$ buffer gas trap (similar to that described in \cite{Cassidy2006}) is used to trap and cool the positrons with a trapping period of approximately 200 ms, resulting in the conversion of the continuous beam into a 5 Hz pulsed beam, with each pulse having FWHM $\sim 50$ ns. The trap contains a rotating wall stage, which, when used in combination with a small amount ($\sim 1\times10^{-6}$ mbar) of CF$_4$ injected into the final stage, greatly increases the positron lifetime in the trap (and therefore the efficiency) by compressing the beam and so decreasing positron annihilation on the trap electrode walls. Despite this, an overall trapping efficiency of only 5\% is recorded, mainly owing to reflection of positrons at the trap inlet. This occurs because the source region magnetic field is considerably lower than the 700 G field surrounding the trap, so positrons with high enough pitch angles can be reflected by the trap inlet electrode as the increased magnetic field has transferred some longitudinal momentum to the transverse direction.\par

\begin{figure}
\centering
\includegraphics[width=\textwidth]{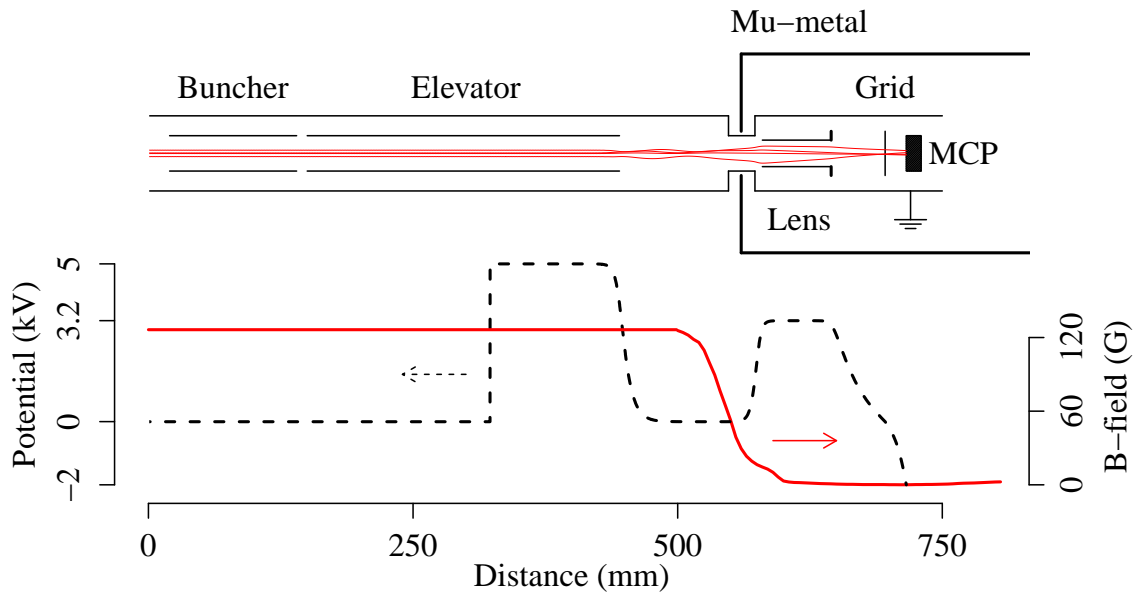}
\caption{Schematic of system showing variation of magnetic field and potential experienced by the positrons.}
\label{fig:exp}
\end{figure}

Upon exiting the trap, the positron pulses are time bunched and then accelerated out of the magnetic field, as described in detail in the following sections. A diagram of the set-up after the trap is shown in Figure \ref{fig:exp}.

\subsection{Time bunching}
\begin{figure}
\centering
\includegraphics[width=0.8\textwidth]{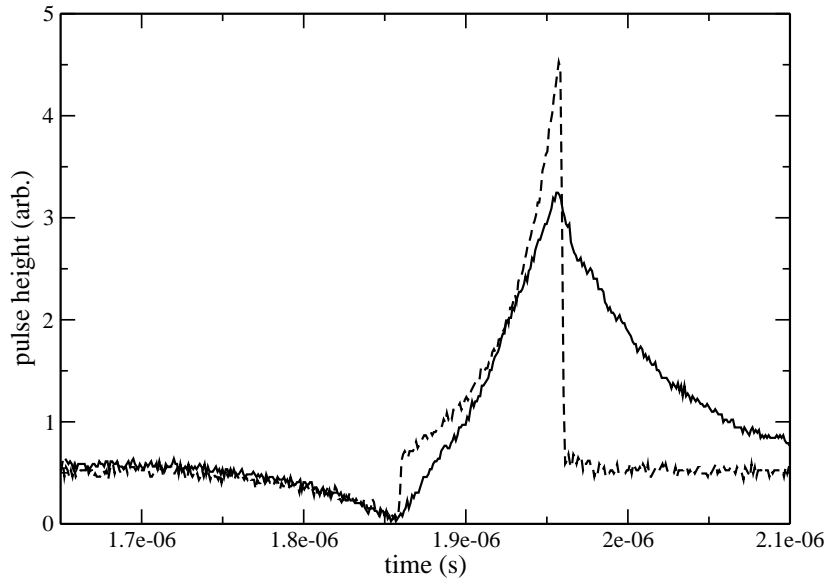}
\caption{Comparison of generated (dashed line) and applied (solid line) bunching potentials.}
\label{fig:buncher_pulse}
\end{figure}

Positron pulses released from the trap have a timing distribution width of approximately 50 ns (FWHM). These pulses are reduced to $\sim$ 1 ns FWHM by passage through a 12 cm bunching tube, to which a time-varying decelerating potential is applied when the positrons enter, and a corresponding accelerating potential is applied as the positrons exit. By assuming that the initial pulse is monoenergetic (it has an energy spread of $\sim 0.06$ eV) and that the acceleration distances are negligible compared to the drift distances inside and after buncher, the decelerating potential $V_d(t)$ can be approximated by:
\begin{eqnarray}
V_d(t) = \frac{E_b}{q_e} - \frac{m_el^2}{2q_e}\left(t_f - t\right)^{-2}
\end{eqnarray}
where $E_b$ is the mean pulse energy, $t_f$ is representative of the bunching timescale, and $l$ is the bunching length scale. The accelerating potential is of a similar form:
\begin{eqnarray}
V_a(t) = \frac{m_el^2}{2q_e}\left(t_f - t\right)^{-2}.
\end{eqnarray}
As the positron bunch can be contained entirely inside the buncher, the accelerating potential can be applied to the same electrode, so that the complete time-varying potential on the electrode can be generated simply by the (amplified) output of an arbitrary waveform generator. This also allows everything before and after the buncher to be held at ground. Although the bunching electrode capacitance is 90 pF and the potential should change rapidly, adequate performance applying the pulse using a single-stage amplifier based around an Analog Devices AD830. A comparison between idealized and applied buncher pulses is shown in Fig \ref{fig:buncher_pulse}.\par

Operation of the buncher (in combination with the elevator) was simulated using SIMION 8.0. The predicted bunch FWHM for a 50 ns wide, gaussian pulse, with a gaussian energy profile with $\sigma = 0.06$ eV is 1.06 ns. Deviations from predicted performance can be attributed to a combination of trigger time jitter, amplifier nonlinearity, and breaking of cylindrical symmetry---the SIMION simulation assumed an axisymmetric geometry. As the MCP used to detect the positrons at the end of the beamline has a single-particle output pulse width of approximately 5 ns the width of the bunched pulse was determined by statistical deconvolution, assuming a gaussian pulse shape. Using this method, a bunch FWHM of $1.26 \pm 0.02$ ns was determined.

\subsection{Elevator}
Following the buncher, the positrons are accelerated using an elevator system comprising a 30 cm cylindrical electrode and high voltage pulsing system. Once the bunch is entirely inside the cylinder, a voltage up to 5000 V is switched in a few ns using a Behkle high voltage MOSFET switch. Owing to the length of the electrode and energy of the positrons, this must be completed inside a $\sim$ 200 ns time window. Upon exiting the electrode, the positrons are accelerated by the new potential difference between the cylinder and the vacuum chamber wall.

\subsection{Extraction from magnetic field}
The accelerated positrons then pass to a region of very low magnetic field, achieved simply by shielding this region with 0.5 mm mumetal with a 16 mm diameter hole for the beam to pass through. A cylindrical electrostatic lens system comprising a single 7 cm focussing element and two split cylinders, refocusses the diverging particles at a tunable distance and allows the position and beam angle to be adjusted independently, prior to entering the final stage of the experiment. SIMION simulations of the trajectories are also shown in Figure \ref{fig:exp}\par

\begin{figure}
\centering
\includegraphics[width=0.8\textwidth]{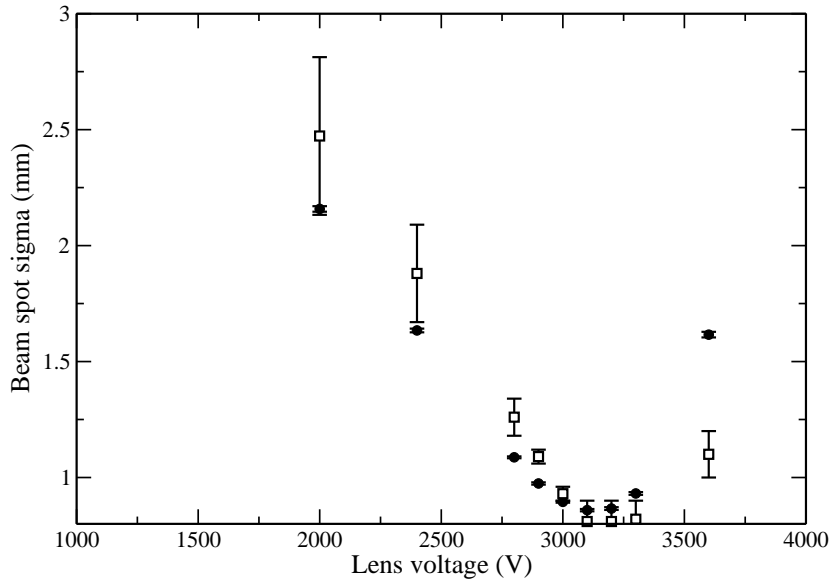}
\caption{Variation of beam spot size with lens voltage($\square$), at an elevator voltage of 4744 V, compared with simulation (\CIRCLE).}
\label{fig:spotsize}
\end{figure}

The performance of this system is extremely sensitive to initial conditions, to the extent that jitter of less than 10 ns in buncher trigger time is detectable at focussed conditions, owing to the time dependence of buncher exit energy. However, by carefully balancing the preceding magnetic field, buncher trigger delay and incoming beam position, it was possible to reproduce reasonably well the expected behaviour of the lens system as simulated using SIMION 8.0. By visually inspecting the beamspot using a position sensitive MCP, and adjusting the aforementioned parameters if distortion was observed, the dependence of spot size on lens voltage shown in Figure \ref{fig:spotsize} was determined. An initial beam spot size of approximately $\sigma = 1$ mm has been measured immediately after the trap.\par

In order to measure the transport efficiency from the trap exit to the lens system exit, a BGO crystal scintillator was used to detect the annihilation photons. An MCP shortly after the exit of the trap provided a position where the input pulse intensity could be measured, with the output pulse being measured at the position of the second MCP. By arranging for the solid angle subtended by the BGO to be the same at both positions, the ratio of the BGO signal provided the transport efficiency. Attempts to use the MCPs themselves led to problems arising from the difference in detection efficiency between the MCPs, as well as the variation of detection efficiency with positron energy, magnetic field, and pulse radius (particularly for long trapping times). Averaging the BGO signal over 512 pulses compensates for the low solid angle subtended by the detector (approximately 0.085 steradians).\par

\begin{figure}
\centering
\includegraphics[width=0.8\textwidth]{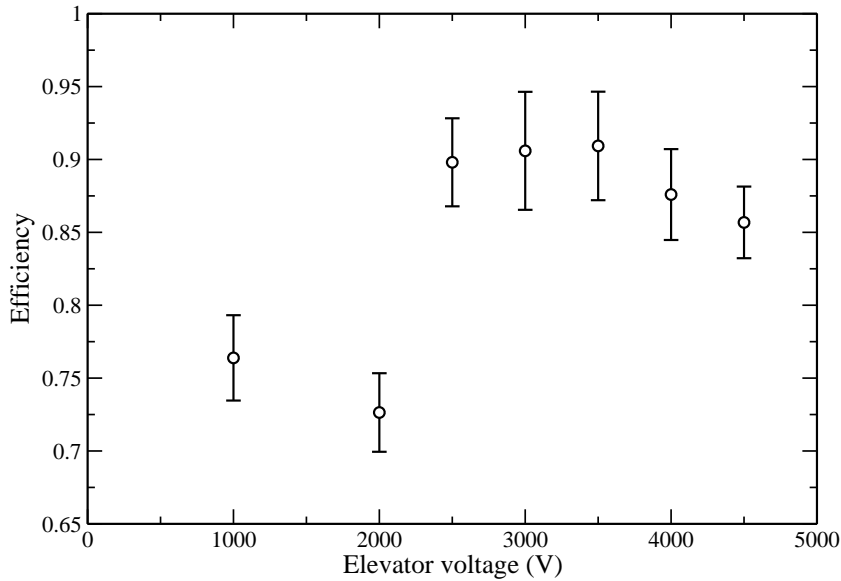}
\caption{Variation of transport efficiency between the trap exit and beamline end (including extraction from the magnetic field) with elevator voltage.}
\label{fig:transport}
\end{figure}

The variation of transport efficiency with elevator voltage is shown in Figure \ref{fig:transport}, and compares well with simulation. 

\section{Conclusions}
The apparatus presented here allows the extraction of positron pulses from a two-stage buffer gas trap to a grounded target region free from magnetic and electric fields. Additionally, the pulse is time compressed to a width on the order of one nanosecond and can be focussed to less than 1 mm. This is achieved in the context of the ongoing research effort into  Ps Rydberg--Stark deceleration \cite{Hogan2011} and Ps 1S--2S spectroscopy at ETH Zurich, however, many experiments may find such high-efficiency extraction combined with good spatial and temporal precision useful.

\ack The authors would like to thank Prof. F. Merkt for useful discussions and technical advice. This work is supported by ETH Zurich and by the SNSF under the grants 200021\_156756 and 206021\_157808.

\section*{References}
\bibliography{bibliography}
\end{document}